\documentstyle[amssymb,prb,aps,manuscript]{revtex}

\begin{document}
\title{Correlation of phonon decay with localized electron spin phase
diffusion}
\author{Y. G. Semenov and K. W. Kim}
\address{Department of Electrical Computer Engineering\\
North Carolina State University, Raleigh, NC 27695-7911}
\maketitle

\begin{abstract}
A spin decoherence mechanism is proposed for localized electrons. The
irregular phonon phase disturbances originated from phonon relaxation can
influence electron spin precession with a net effect of spin phase decay. A
quantitative analysis demonstrates relatively high efficiency of this
mechanism in the low temperature and low magnetic field regime compared to
the spin-flip processes.
\end{abstract}



\section{Introduction}

Recently, much attention has been devoted to electron spin
relaxation in quantum dots (QDs) since they provide a natural
candidate for the qubits in quantum computing. A typical approach
to this problem is to calculate the spin transition probability
associated with the spin-flip processes, i.e., longitudinal spin
relaxation. However, quantum computing is qualitatively limited by
the processes that result in the destruction of electron spin
phase coherence. For example, phase diffusion of localized
electron spin can be characterized by relaxation mechanisms that
are not related to spin-flip
processes under certain conditions. Hence, further investigation of {\em %
transversal} (or phase) relaxation $T_{2}$ is crucial for accurate
understanding.

One such mechanism was proposed in Ref.~\onlinecite{SemKim03} where a random
change of spin precession and subsequent spin phase diffusion is associated
with the transitions between electronic quantum states with different $g$
factors. Although generally efficient, this process is frozen out at low
temperatures due to its phonon-mediated nature and the direct spin-flip is
expected to be the dominant mechanism of phase relaxation. However, the
spin-flip relaxation reveals a very strong (4th to 5th power) dependence on
the magnetic field,~\cite{KhaetNaz01,GlavinKim} becoming rather ineffective
at low fields. Hence, it is necessary to explore other potential sources of
decoherence, particularly in the low field and low temperature regime. In
this work, we show that the spin-phonon interaction, which heretofore was
considered mainly with respect to the resonant processes, can provide such a
mechanism if a finite phonon damping is taken into account.

Our analysis is based on the representation of spin-phonon interaction in
terms of fluctuating effective magnetic field $\overrightarrow{\Omega }$ (in
units of energy) acting on the electron spin $\overrightarrow{s}$. This
field is assumed to be composed of additive contributions $\overrightarrow{
\Omega }_{p}$ from each phonon $p=\{\overrightarrow{q},\varkappa \}$ with a
wave vector $\overrightarrow{q}$ and polarization $\varkappa $, i.e., $%
\overrightarrow{\Omega }= \Sigma_{p} \overrightarrow{\Omega}_{p}$.
For the moment, let us focus on a single phonon contribution.
Then, in the frame of reference rotating with the Zeeman
frequency, the electron spin performs precession around the small
$\overrightarrow{\Omega }_{p}$, which oscillates with a phonon
frequency $\omega _{p}$. No alteration in the electron spin phase
occurs due to such a harmonic perturbation with a possible
exception of spin phase shift $\Delta \phi _{0}$ acquired at the
initial period of interaction $0<t<2\pi /\omega _{p}$ due to a
random phonon phase $\theta _{p} $.~\cite{Klemens}

A different situation can be realized when a phonon harmonic oscillation is
interrupted and resumes at a series of instant times $t_{1i}$ and $t_{2i}$ ($%
i=0,1,...$), respectively. The reason of such phonon fluctuations can be
lattice anharmonicity, phonon scattering at the impurities or lattice
defects, etc. These irregular phonon perturbations affect the electron spin
precession resulting in the phase shift $\Delta \phi _{i}$ at each interval
of time $t_{2i}-t_{1i}$. Subsequently, the net effect of spin phase change $%
\phi _{p}(t)$ due to a phonon mode $p$ can be expressed as $\phi
_{p}(t)=\Sigma _{i}\Delta \phi _{i}$, ($t_{2i}<t$).

Note that for a large number of small changes $\Delta \phi _{i}$, their
total effect can be described by a diffusion equation. Its solution leads to
an exponential decay of electron spin phase with a relaxation rate $%
T_{p}^{-1}=\frac{1}{2}\left\langle \Delta \phi _{i}^{2}\right\rangle \tau
_{p}^{-1}$, where $\tau _{p}$ is the mean time between sequential instants $%
t_{1i}$ (or $t_{2i}$).\cite{Sem03} To estimate the spin phase change $\Delta
\phi _{i}$ caused by a phonon perturbation during the $t_{2i}-t_{1i}$, it is
helpful to recognize that a single oscillator influence does not change a
spin phase during its full period $\Delta t_{p}=2\pi /\omega _{p}$ as well
as for any $n$ integer periods $n2\pi /\omega _{p}$. Hence, $\Delta \phi
_{i} $ can be approximated as a spin rotation $\Omega _{p}\Delta t_{p}$ in
an effective field $\overrightarrow{\Omega }_{p}$ independently on duration $%
t_{2i}-t_{1i} $. With the mean value $\left\langle \Delta \phi
_{i}\right\rangle $ on the order of $\Omega_{p}/ \omega_{p}$, one can expect
$T_{p}^{-1} \sim \tau_{p}^{-1} \Omega_{p}^{2}/\omega _{p}^{2}$ for the
phonon mode $p$ and $T_{2}^{-1}\sim \Sigma_{p} N_{p} \tau_{p}^{-1}
\Omega_{p}^{2}/\omega _{p}^{2}$ when the contributions of all phonons (with
the population factor $N_{p}$) are taken into account.

The qualitative consideration provided above shows that electron spin phase
relaxation can be strongly affected by phonon phase damping of any origin
such as phonon decay. Moreover, since this mechanism does not involve energy
exchange, only the longitudinal (with respect to the external magnetic field
$\overrightarrow{B}$) component $\Omega _{z}$ of the effective fluctuating
field is relevant to our case. These characteristics qualitatively
distinguish the mechanism under consideration from other processes, most of
which are determined by fluctuations of transversal components $\Omega _{x}$
and $\Omega _{y}$ at the resonant frequency with the Zeeman splitting.

\section{Theoretical Model}

For a detailed quantitative analysis of the proposed mechanism, let us start
with the spin-phonon interaction operator
\begin{equation}
H_{s-ph}=\overrightarrow{\Omega }\vec{s},  \label{eq1}
\end{equation}%
where the $\alpha $-th component ($\alpha =x,y,z$) of the fluctuating field
takes a form linear in the creation and annihilation operators $a_{p}^{\dag
} $ and $a_{p}$ of the phonon mode $p$ [$-p\equiv \{-\overrightarrow{q}%
,\varkappa \}$]; i.e.,
\begin{equation}
\Omega _{\alpha }=\sum_{p}V_{\alpha }^{p}Q_{p}\equiv \sum_{p}V_{\alpha
}^{p}\left( a_{p}^{\dag }-a_{-p}\right)  \label{eq2}
\end{equation}%
with a matrix element $V_{\alpha }^{p}$ of the spin-phonon interaction. The
specific form of $V_{\alpha }^{p}$ will be discussed later.

Now we focus on the spin evolution caused by random fluctuations of $\Omega
_{\alpha }$. Obviously electron spin follows each of such fluctuations that
result in its irregular behavior at the time scale $\tau _{c}$ of the $%
\Omega _{\alpha }$ fluctuations. Actually a random single spin fluctuation
associated with each phonon scattering is expected to be very small and
drops out of the problem; instead, the total result of these small
fluctuations averaged over the time scale $\Delta t$ ($\tau _{c}\ll \Delta
t\ll T_{2}$) is the subject of our investigation. The time evolution of mean
spin value $\vec{s}$ can be described by the quantum kinetic equation~\cite%
{Sem03} in the case of anisotropic medium and interaction $H_{s-ph}$ [Eq. (%
\ref{eq1})]%
\begin{equation}
\frac{d}{dt}\vec{s}(t)=\vec{\omega}\times \vec{s}(t)-{\bf \Gamma }\left[
\vec{s}(t)-\vec{s}_{0}\right] ,  \label{eq3}
\end{equation}%
where $\vec{\omega}$ is an effective field with components $\omega
_{i}=\sum_{j}g_{ij}\mu _{B}B_{j}$.\cite{comm2} As usual, $g_{ij}$ are the
components of $g$ tensor, the subscripts $i$ and $j$ relate to the
crystalline coordinate system, $\mu _{B}$ is the Bohr magneton, $%
\left\langle \ldots \right\rangle =Tr\{e^{-H_{d}/T}\ldots \}/Tre^{-H_{d}/T}$
where $H_{d}$ is the Hamiltonian of the dissipative subsystem (lattice
vibrations in our case), and $T$ is the temperature. $\vec{\omega}$ and $T$
are expressed in units of energy. The matrix ${\bf \Gamma }$ of relaxation
coefficients is composed of Fourier transformed correlation functions
\begin{equation}
\gamma _{\mu \nu }\equiv \gamma _{\mu \nu }\left( \omega \right)
=\left\langle \Omega _{\mu }\left( \tau \right) \Omega _{\nu }\right\rangle
_{\omega }=\frac{1}{2\pi }\int\limits_{-\infty }^{\infty }\left\langle
\Omega _{\mu }\left( \tau \right) \Omega _{\nu }\right\rangle e^{i\omega
\tau }d\tau  \label{eq3a}
\end{equation}%
with $\Omega _{\mu }\left( \tau \right) =\exp (iH_{d}\tau )\Omega _{\mu
}\exp (-iH_{d}\tau )$. It has a canonical form in the frame of references $%
\widehat{x}$, $\widehat{y}$, $\widehat{z}$ with $\widehat{z}$ directed along
$\vec{\omega}$ (so that $\mu $, $\nu =x$, $y$, $z$). With a provision that
the correlation functions are symmetrical, $\gamma _{\mu \nu }\left( \omega
\right) =\gamma _{\nu \mu }\left( \omega \right) $, the matrix $\Gamma $ has
a simpler form: $\Gamma _{xx}=\pi \left( \gamma _{zz}^{0}+n\gamma
_{yy}\right) $, $\Gamma _{yy}=\pi \left( \gamma _{zz}^{0}+n\gamma
_{xx}\right) $, $\Gamma _{zz}=\pi n(\gamma _{xx}+\gamma _{yy})$, $\Gamma
_{\mu \nu }=-\pi n\gamma _{\mu \nu }$ , ($\mu \neq \nu $), where $\gamma
_{zz}^{0}=\gamma _{zz}(0)$, $n\equiv n(\omega )=\left( 1+e^{\omega
/T}\right)/2$, $\vec{s}_{0}=-{\frac{1}{2}}\widehat{z}\tanh (\omega/2T)$, $%
\omega =\omega _{z}=\left( \sum_{i}\omega _{i}^{2}\right) ^{1/2}$.

One can see that the coefficients $\Gamma_{xx}$ and $\Gamma_{yy}$
responsible for transversal relaxation consist of two parts, $T_{2,\omega
}^{-1}=\pi n\gamma _{yy}$ (or $\pi n\gamma _{xx}$) and $T_{2,0}^{-1}=\pi
\gamma _{zz}^{0}$. Comparison with the longitudinal relaxation coefficient $%
\Gamma _{zz}$ shows that the term $T_{2,\omega }^{-1}$ stems from the
contribution of spin-flip processes involving energy exchange between the
Zeeman and phonon reservoirs. Since the longitudinal relaxation has been the
subject of a number of recent studies,~\cite%
{KhaetNaz01,GlavinKim,Tahan,ErlNazFal} we focus on the analysis of $%
T_{2,0}^{-1}$ term.

The correlation function Fourier image [Eq.~(\ref{eq3a})] of the effective
field $\overrightarrow{\Omega }$ is expressed in terms of the phonon
operators according to Eq.~(\ref{eq2}). In turn, the Fourier image of phonon
correlation functions $\varphi _{p}(\omega )=\left\langle Q_{p}(\tau
)Q_{-p}\right\rangle _{\omega }$ is Lorentzian-like since the corresponding
Green function satisfies the equation $G_{p}(\omega )=(\omega _{p}/\pi
)[\omega ^{2}-\omega _{p}^{2}-2\omega _{p}M_{p}(\omega )]^{-1}$, where the
"mass" operator $M_{p}(\omega )$ depends on the phonon interaction (see, for
example, Ref.~\onlinecite{Zubarev}). In the most general case, this
correlation function takes the form~\cite{Pathak}
\begin{equation}
\varphi _{p}(\omega )=\frac{1}{\pi }\frac{\left( 2N_{p}+1\right) \Gamma
_{p}(\omega )}{\left( \omega ^{2}-\omega _{p}^{2}\right) ^{2}/\omega
_{p}^{2}+\Gamma _{p}^{2}(\omega )},  \label{eq3b}
\end{equation}%
where $\Gamma _{p}(\omega )=%
\mathop{\rm Im}%
M_{p}(\omega )$ depends on the specific mechanism of phonon scattering. In
such a manner, $\Gamma _{p}(\omega )$ is a function of temperature due to
the anharmonicity of the third and fourth order; furthermore, there are
contributions by other sources of phonon scattering (point defects,
isotopes, dislocations, crystal boundaries and interfaces) that reveal
different dependencies on $\omega $ and $\omega _{p}$. Hence, evaluation of
the relaxation coefficients becomes too complicate to be approached
analytically. Instead, to proceed further, we utilize the phonon relaxation
time that can be extracted from the thermal conductivity measurements (see
Ref.~\onlinecite{HanKlemens} and the references therein). An expression
appropriate for the correlation function Fourier image $\gamma _{\mu \nu
}\left( \omega \right) $ was derived in Ref.~\onlinecite{Sem03} in the
relaxation time approximation. For our particular case of $\omega =0$ and $%
\mu =\nu =z$, it can be reduced to
\begin{equation}
\gamma _{zz}^{0}=\sum_{p_{1},p_{2}}V_{z}^{p_{1}}V_{z}^{p_{2}}\left\langle
Q_{p_{1}}Q_{p_{2}}\right\rangle \frac{1}{\pi }\frac{\tau _{p_{1}}^{-1}}{%
\omega _{p_{1}}^{2}+\tau _{p_{1}}^{-2}},  \label{eq4}
\end{equation}%
where $\tau _{p}=1/\Gamma _{p}(\omega _{p})$ is the relaxation time of
phonon mode $p$ (i.e., phonon lifetime). In most cases, one can assume $%
\omega _{p}\gg \tau _{p}^{-1}$ and neglect the second term in the
denominator of Eq.~(\ref{eq4}). Then, along with the definition of the
operator $Q_{p}$ [see Eq.~(\ref{eq2})], one can express the non-resonant
phonon contribution to the transversal spin relaxation rate in the form
\begin{equation}
T_{2,0}^{-1}=\sum_{p}\left\vert V_{z}^{p}\right\vert ^{2}\left(
2N_{p}+1\right) \frac{\tau _{p}^{-1}}{\omega _{p}^{2}},  \label{eq5}
\end{equation}%
which is in accordance with the qualitative analysis discussed earlier in
this paper. The phonon population factor $N_{p}$ is given as $\left[ \exp
\left( \omega _{p}/T\right) -1\right] ^{-1}$.

Equation~(\ref{eq5}) is the starting point of our investigation on
the proposed spin relaxation mechanism.  However, this still
requires the detailed knowledge of the phonon dispersion
$\omega_{p}$ and the relaxation time $\tau _{p}$ for each phonon
mode $p$. By taking into account the conditions frequently
encountered in quantum computation utilizing semiconductor QDs, we
restrict our consideration to the case when the radius $a_{0}$ of
the electron state is much larger than the lattice constant and
the temperature is sufficiently low. Since the spin-phonon
interaction matrix $V_{z}^{p}$ is significant only for the phonon
wave vector $q \lesssim 1 /a_{0}$, a large $a_{0}$ essentially
limits the summation of Eq.~(\ref{eq5}) to long wavelength
phonons.
Subsequently $\tau _{p}$, which is a complex function of the
temperature and phonon frequencies,~\cite{Carruthers} can be
considered in the long wavelength limit. Moreover, at low enough
temperatures $T\lesssim T_{bs}$ ($T_{bs}\approx 10$ K in the case
of Ref.~\onlinecite{Carruthers}), only one term originating from
the boundary scattering survives for phonon
relaxation.~\cite{HanKlemens} Since this mechanism is insensitive
to the temperature as explained by Ref.~\onlinecite{HanKlemens},
it is adequate to assume a constant phonon relaxation time $\tau
_{p}\simeq \tau _{ph}$ for long wavelength phonons at $T\lesssim
T_{bs}$.  This permits us to avoid the problems associated with
the complex dependence of $\omega_{p}$ and $\tau _{p}$, which can
be very specific for each particular sample.

\subsection{Effect of ${\protect\small {g}}$-factor fluctuation}

To evaluate $V_{z}^{p}$, we consider the spin-lattice interaction via phonon
modulation of $g$ factor. In general, the spin-lattice interaction
Hamiltonian can be written in terms of the tensor $A_{ijkl}$:\cite%
{Roth,Koloskova,GlavinKim}
\begin{equation}
H_{s-ph}=\mathop{\displaystyle\sum}\limits_{ijkl}A_{ijkl}\mu _{B}B_{i}s_{j}%
\overline{u_{kl}},  \label{eq7}
\end{equation}%
where $\overline{u_{kl}}$ is the strain tensor $u_{kl}$ averaged over the
electron ground state $\left\vert g\right\rangle =\psi _{g}(\overrightarrow{r%
})$: $\overline{u_{kl}}=\left\langle g\left\vert u_{kl}\right\vert
g\right\rangle $. By way of important example, we consider a $z$-directed
magnetic field and a localized electron with the axial symmetry with respect
to the $z$-axis. This reduces Eq.~(\ref{eq7}) to the form of Eq.~(\ref{eq1})
with $\Omega _{z}=\left[ \left( A_{33}-A_{31}\right) \overline{u_{zz}}+A_{31}%
\overline{\Delta }\right] \mu B$; here, $\Delta $ denotes the dilatation $%
\Delta =\Sigma _{i}u_{ii}$ and the Voigt notation is adopted ($%
A_{33}=A_{zzzz}$, $A_{31}=A_{zzxx}$, $A_{66}=A_{xyxy}$). Then, the matrix
element of the spin-phonon interaction takes the expression
\begin{equation}
V_{z}^{p}=i\left( \frac{\hbar }{2\rho V\omega _{p}}\right) ^{1/2}\left[
\left( A_{33}-A_{31}\right) e_{z}^{p}q_{z}+\delta _{\varkappa ,L}A_{31}q%
\right] \Phi \left( \overrightarrow{q}\right) \mu _{B}B,  \label{eq8}
\end{equation}%
where $\rho $ is the mass density of the crystal, $V$ is the
volume of the sample structure, $\overrightarrow{e}^{p}$ the
polarization vector of the phonon mode $p$, $\varkappa  =L,T$, and
$\Phi \left( \overrightarrow{q}\right) =\left\langle g\right\vert
e^{i\overrightarrow{q}\cdot \overrightarrow{r}}\left\vert
g\right\rangle $. The spin-lattice relaxation rate in
Eq.~(\ref{eq5}) can be calculated by treating the phonon modes
based on the isotropic elastic continuum model with the
longitudinal and transverse sound velocities $c_{L}$ and $c_{T}$.
Assuming the axial symmetry for the local electron center, i.e.
$\Phi \left( \overrightarrow{q}\right) =\Phi \left( x,z\right) $
($x=qa_{0}/2 $, $z=q_{z}/q$, the parameter $a_{0}$ represents the
electron state radius as mentioned before), one can obtain
\begin{equation}
T_{2,0}^{-1}=\tau _{ph}^{-1}\xi (B)\displaystyle\int_{0}^{x_{\max }}x\frac{%
\tau _{p}^{-1}}{\tau _{ph}^{-1}}\left[ \coth \left( \frac{T_{T}^{eff}}{T}%
x\right) F_{T}\left( x\right) +\frac{c_{T}^{3}}{c_{L}^{3}}F_{L}\left(
x\right) \coth \left( \frac{T_{L}^{eff}}{T}x\right) \right] dx,  \label{eq9}
\end{equation}%
\begin{eqnarray}
\xi (B) &=&\frac{\left( A_{33}-A_{31}\right) ^{2}\mu _{B}^{2}B^{2}}{2\pi
^{2}\hbar \rho c_{T}^{3}a_{0}^{2}},  \nonumber \\
F_{L}\left( x\right)  &=&\displaystyle\int_{-1}^{1}\left( z^{2}+\zeta
\right) ^{2}\Phi ^{2}\left( x,z\right) dz,  \label{eq10} \\
F_{T}\left( x\right)  &=&\displaystyle\int_{-1}^{1}z^{2}(1-z^{2})\Phi
^{2}\left( x,z\right) dz,  \nonumber
\end{eqnarray}%
where $\tau _{ph}^{-1}$ is an average phonon relaxation rate, $T_{\varkappa
}^{eff}=\hbar c_{\varkappa }/k_{B}a_{0}$ is the effective temperature, and $%
\zeta \equiv A_{31}/\left( A_{33}-A_{31}\right) =-1/3$ if one assumes that
the strain induced part of the effective $g$-tensor $\widetilde{g}%
_{ij}=\Sigma _{k,l}A_{ijkl}u_{kl}$ is characterized by zero trace, i.e., $%
A_{33}+2A_{31}=0$. When $a_{0}$ is much larger than the lattice
constant, the upper limit $x_{\max }$ in the integral of
Eq.~(\ref{eq9}) may be taken to infinity since $\Phi
(\overrightarrow{q})$ restricts the actual phonon wave vectors to
$q\lesssim 1/a_{0}$ as discussed above.

Let us evaluate spin relaxation of a shallow donor with an
effective Bohr radius $a_{B}$ (=$a_{0}$) and $\Phi \left(
x,z\right) =\left( 1+x^{2}\right) ^{-2}$. Utilizing the constant
phonon relaxation time approximation $\tau _{p}\simeq \tau _{ph}$
for $T\lesssim T_{bs}$, the integral in Eq.~(\ref{eq9}) can be
evaluated analytically
\begin{equation}
T_{2,0}^{-1}=\frac{2\xi (B)\tau _{ph}^{-1}}{45}\left( \sqrt{1+\frac{T^{2}}{%
T_{\perp }^{2}}}+\frac{2c_{T}^{3}}{3c_{L}^{3}}\sqrt{1+\frac{T^{2}}{%
T_{\shortparallel }^{2}}}\right) ,  \label{eq11}
\end{equation}%
where $T_{\shortparallel (\perp )}=\left( 16/15\pi \right)
T_{L(T)}^{eff}$. Note that Eq.~(\ref{eq11}) is obtained with $
\overrightarrow{B}\parallel \lbrack 001]$.  In the case of cubic
symmetry [where only two constants $A_{66}$ and $A_{33}=-2A_{31}$
in Eq.~(\ref{eq7}) describe the effect of spin-phonon coupling],
an expression $T_{2,0}^{-1}$ for an arbitrarily directed
$\overrightarrow{B} $ can be obtained in terms of the direction
cosines $l=B_{x}/B$, $m=B_{y}/B$, $n=B_{z}/B$. Our calculations
show that this is achieved by multiplying the factor
\begin{equation}
f(\overrightarrow{B}/B)=1+\left( \frac{4}{9}\frac{A_{66}^{2}}{A_{33}^{2}}%
-1\right) P,  \label{eq11b}
\end{equation}%
to Eq.~(\ref{eq11}); $ P=3(l^{2}m^{2}+m^{2}n^{2}+n^{2}l^{2})$,
$0\leq P\leq 1$. One can see that the angular dependence of our
mechanism does not result in zero relaxation under any direction
of $ \overrightarrow{B}$. Moreover, the directions along the
principal axes ($[001]$, etc.) can result in maximal relaxation,
while the same directions sometimes forbid the spin-flip
processes.\cite{Roth,Hasegawa}

As an example, we consider a Phosphorus shallow donor in Si with
$a_{B}=1.8$~nm. The phonon relaxation time can be extracted from
the low temperature measurements of Si thermal
resistivity~\cite{GlassSlack} in terms of the theory developed in
Refs.~\onlinecite{Callaway} and \onlinecite{HanKlemens} ($\tau
_{ph}=2.4\times 10^{-8}$ s). The spin-phonon coupling constants
were estimated in the works of Refs.~\onlinecite{Roth} and
\onlinecite{Hasegawa}. However, we believe that direct
determination of coupling constants by means of EPR measurements
of Si:P under an applied stress gives more reliable data. A
corresponding experiment was performed in Ref.~\onlinecite
{WilsonFeher}, where the constant $A_{66}=0.44$ was found.
Similarly, our estimation obtained $A_{33}=0.31$ and
$A_{31}=-0.155$ that gives $ T_{2,0}^{-1}=1.3\times 10^{-4}$
s$^{-1}$ at the magnetic field of 1 T and low temperatures $T\ll
T_{\shortparallel (\perp )}\simeq 10$ K.

In another important case of a Si shallow donor in Al$_{0.4}$Ga$_{0.6}$As,
the data on EPR under a uniaxial stress\cite{Glaser} provide rather strong
spin-phonon constants of $A_{33}=19.6$ and $A_{31}=-9.8$. This gives the
estimation $T_{2,0}^{-1}=6.1 \times 10^{-2}$ s$^{-1}$ and $6.1 \times
10^{-4} $ s$^{-1}$ for the magnetic fields of 1 T and 0.1 T, respectively,
at $T=4$~K under the assumption that phonon lifetimes are identical in these
crystals.

Similar calculations can be performed for an electron localized in a QD of $%
L_{xy}=2a_{0}$ in the lateral width and $L_{w}=\epsilon L_{xy}$ in the
thickness. Under the condition $\epsilon \lesssim 0.1$, an approximate
formula takes the form
\begin{equation}
T_{2,0}^{-1}=\xi (B)\tau _{ph}^{-1}\left( \sum_{i=L,T}b_{i}\sqrt{%
c_{i}^{2}+d_{i}^{2}\frac{T^{2}}{T_{i}^{2}}}\right) ,  \label{eq12}
\end{equation}%
where the fitting coefficients are $b_{T}=1$, $b_{L} = c_{T}^{3}/c_{L}^{3}$,
$c_{T}=0.33-1.27\epsilon ^{2}$, $d_{T} = 0.35-0.395\epsilon ^{2}$, $c_{L} =
0.97 - 28.5\epsilon ^{2}$, and $d_{L} = 0.40 - 3.76\epsilon ^{2}$.

Let us compare, as an example, spin phase relaxation caused by the phonon
decay [Eq.~(\ref{eq12})] with the spin-flip admixture mechanism (Ref.~%
\onlinecite {KhaetNaz01}) in a GaAs QD with $L_{w}=3$~nm and $L_{xy}=25$~nm,
assuming $\tau _{ph}=2.4\times 10^{-8}$ s and $A_{33}=19.6$. For the
relatively strong magnetic field of 1 T and $T=4$~K, our mechanism and the
spin-flip mechanism give $T_{2,0}^{-1}\approx 0.1$~s$^{-1}$ and $\frac{1}{2}%
T_{1}^{-1}=T_{2,\omega }^{-1}=10$~s$^{-1}$, respectively, while for $B=0.1$%
~T both mechanisms predict almost the same rate of $\approx 10^{-3}$~s$^{-1}$%
. In lower magnetic fields, our mechanism prevails.

\subsection{Effect of hyperfine constant modulation}

The $g$-factor modulation described in Eq.~(\ref{eq7}) is not the only
possible mechanism of spin-phonon interaction. For an alternative process,
let us consider the hyperfine interaction (HFI) of localized electrons with
the nuclei:
\begin{equation}
H_{hf}=a_{hf}\sum_{j}\left\vert \psi (\overrightarrow{r}_{j})\right\vert ^{2}%
\overrightarrow{I}_{j}\overrightarrow{s},  \label{eq13}
\end{equation}%
where $a_{hf}$ is the HFI constant and $\overrightarrow{I}_{j}$ is the
nuclear spin situated at site $j$ with the position $\overrightarrow{r}_{j}$%
. Lattice vibrations near the nuclear equilibrium positions can lead to
effective field fluctuations and, subsequently, the spin-phonon interaction.
Taking into account the long wavelength phonons with respect to the mean
internuclear distance $\approx n_{i}^{-1/3}$ ($n_{i}$ is the nuclear spin
concentration), the main part of this interaction for a typical nuclear spin
configuration can be represented as in Eq.~(\ref{eq1}) with
\begin{equation}
\overrightarrow{\Omega }=\widehat{n}\sqrt{I(I+1)n_{i}/V_{QD}}a_{hf}\overline{%
\Delta }.  \label{eq14}
\end{equation}%
Here, the unit vector $\widehat{n}$ is directed along the effective nuclear
field defined by Eq.~(\ref{eq13}) and $V_{QD}=\left( \int \left\vert \psi (%
\overrightarrow{r})\right\vert ^{4}d^{3}\overrightarrow{r}\right) ^{-1}$.
Calculation of the phase relaxation rate for the case of a shallow donor
results in the expression, which is similar to Eq.~(\ref{eq11}),
\begin{equation}
T_{2,0}^{-1}=\frac{\xi _{hf}\tau _{ph}^{-1}}{3}\sqrt{1+\frac{T^{2}}{%
T_{\shortparallel }^{2}}},  \label{eq15}
\end{equation}%
where the parameter%
\begin{equation}
\xi _{hf}=\frac{I(I+1)n_{i}a_{hf}^{2}}{6\pi ^{2}\hbar \rho
V_{QD}c_{L}^{3}a_{0}^{2}}  \label{eq16}
\end{equation}%
is independent on the magnetic field. In the case of an electron localized
in a QD, one can find the approximate rate through an analogy with Eq.~(\ref%
{eq12}):
\begin{equation}
T_{2,0}^{-1}=\xi _{hf}\tau _{ph}^{-1}\sqrt{c_{hf}^{2}+d_{hf}^{2}\frac{T^{2}}{%
T_{L}^{2}}},  \label{eq17}
\end{equation}%
where $c_{hf}=3.7-68\epsilon ^{2}$, $d_{hf}=2.7-9.8\epsilon ^{2}$, and $%
\epsilon \lesssim 0.1$. Numerical estimations provided for a donor in Si and
GaAs in terms of Eq.~(\ref{eq15}) indicate inefficiency of this mechanism
with a very long relaxation time (about 10$^{14}$~s and 10$^{8}$~s,
respectively). Hence, this mechanism can be neglected in most cases.

\subsection{Two phonon process}

So far, we primarily considered the influence of phonon decay on
spin phase relaxation via linear spin-phonon interaction as given
in Eq.~(\ref{eq7}). Namely, the effect of phonon scattering with
an electron spin on phonon relaxation has not been considered
(i.e., electron spin-induced phonon decay). The Hamiltonian of
this process can be derived in terms of spin-two-phonon
interaction $H_{s-ph}^{(2)}=\sum D_{ijklmn}\mu
_{B}B_{i}s_{j}u_{kl}u_{mn}$ with the spin-phonon coupling constants $%
D_{ijklmn}$. Now the fluctuating effective field takes the form $\Omega
_{\alpha }=\sum_{p,p^{\prime }}W_{\alpha }^{p,p^{\prime }}Q_{p}Q_{p^{\prime
}}$ ($W_{\alpha }^{p,p^{\prime }}$ are the matrix elements of $%
H_{s-ph}^{(2)} $), so the correlation function Fourier image
$\gamma _{\mu \nu }\left( \omega \right) $ [Eq.~(\ref{eq3a})] is
expressed in terms of phonon correlation functions $\left\langle
(Q_{p_{1}}Q_{p_{2}})\left( \tau \right)
Q_{p_{3}}Q_{p_{4}}\right\rangle _{\omega }$. Its calculation
performed in a harmonic approximation leads to a simple expression
$\delta (\omega _{p_{1}}-\omega _{p_{2}})(\delta
_{p_{1},p_{3}}\delta _{p_{4},p_{2}}+\delta _{p_{2},p_{3}}\delta
_{p_{4},p_{1}})(2N_{p_{1}}N_{p_{2}}+N_{p_{1}}+N_{p_{2}}) $.
Substituting this function for $\gamma _{\mu \nu }\left( \omega
\right) $ and a parameter $D$ for the dominant contribution among
the coupling constants $D_{ijklmn}$, the spin phase relaxation
rate for the two-phonon process is given at low temperatures
($T<\hbar c_{T}/k_{B}a_{0}$) approximately as
\begin{equation}
T_{2,0}^{-1}=\frac{\mu _{B}^{2}B^{2}D^{2}}{21\rho ^{2}c_{T}^{3}}\left( \frac{%
k_{B}T}{\hbar c_{T}}\right) ^{7}.  \label{eq18}
\end{equation}%
Parameter $D$ can be estimated as $D=3(g-2)C^{2}/E_{g}^{2}$ ($g$,
$C$, and $E_{g}$ are the electron $g$ factor, deformation
potential and energy gap).~\cite{SemKim03}  Numerical evaluation
of Eq.~(\ref{eq18}) at low temperatures ($T=4$ K) predicts a long
relaxation time. In the case of GaAs at $B=1$~T, one can find $
T_{2,0}\approx 3\times 10^{5}$~s, which is too long to be of any
experimental or practical interest.

\section{Discussion}

To illustrate the significance of the mechanism under
consideration, let us briefly survey the most important spin
decoherence mechanisms reported in the literature: the HFI and
spin-lattice interactions. In the presence of the HFI, an electron
spin performs precession around the sum of the external magnetic
field $\overrightarrow{B}$ and the effective field
$\overrightarrow{B}_{hf}$ caused by the HFI. Dispersion of
$\overrightarrow{B}_{hf}$ over an ensemble of QDs results in a
relatively fast electron spin phase diffusion (see
Refs.~\onlinecite{Merkulov} and \onlinecite{SemKim02}); however,
it causes only a partial dephasing ($<67\%$) and can be
essentially eliminated as $B>>B_{hf}$ ($B_{hf}<1 $ G for typical
Si QDs).

In the case of a single electron in a QD, the electron spin can change its
phase through the HFI since the nuclei also perform precession around the
effective field caused by the electron spin. This field proportional to $%
\left\vert \psi (\overrightarrow{r}_{j})\right\vert ^{2}$ [see Eq.~(\ref%
{eq13})] is inhomogeneous over the QD volume, which distorts the mutual
correlation of nuclei spin configuration and subsequently causes an
alteration in the direction and strength of $\overrightarrow{B}_{hf}$.~\cite%
{Merkulov,KhaetskiLoss} However, this relaxation is rather long and can be
suppressed if $B>>B_{hf}$. In addition, it can be further reduced in the
case of full nuclei spin polarization~\cite{KhaetskiLoss} and/or isotope
purification. Hence, the spin-lattice (i.e., phonon) interaction provides
the most fundamental and unavoidable source of electron spin decoherence.

Among the spin-lattice interaction mechanisms, the phonon-mediated
transitions between the ground and excited states modulate the
precession velocity leading to very effective
decoherence,~\cite{SemKim03} when their energy separations are
small enough. However, under the assumption $ k_{B}T\ll \delta
_{0}$ this relaxation is reduced as $\exp (-\delta _{0}/k_{B}T)$.
Thus, the spin-flip processes and the phonon-decay induced
mechanism considered in this paper provide the main contributions
at low temperatures. Moreover, these two mechanisms differ in the
magnetic field dependence. When the magnetic field decreases, the
spin-flip process yields to spin phase diffusion induced by phonon
relaxation as mentioned above. The estimated magnetic field
strength for this cross-over (e.g., $\lesssim$ 0.1 T) is well
within the range of practical importance.

\section{Conclusion}

We considered spin phase diffusion of a localized electron through
anharmonic phonon disturbances. In contrast to the spin-flip
process where only the resonant (with the Zeeman energy) phonons
are relevant, electron spin phase acquires random shifts when
relaxation of any (resonant or nonresonant) phonon occurs. A
quantitative analysis shows that the considered phase relaxation
reveals a relatively weak dependence on the magnetic field
strength and the temperature compared to the direct spin
relaxation processes or other mechanisms that involve the excited
electron states. In addition, a specific dependence on the
magnetic field direction [Eq.~(\ref{eq11b})] is attributed to this
mechanism. Thus, one can expect that at low temperatures and
magnetic fields the spin phase diffusion mediated by the phonon
relaxation can become dominant over the spin-flip processes. As
for quantitative estimation of the relaxation rate, the decisive
role belongs to the phonon lifetime $\tau_{ph}$. In the present
study, we estimated $\tau_{ph}$ from the experiments conducted in
bulk Si. It is not apparent if this estimation is applicable to
the case of QDs. Moreover, the phonon lifetime may be a function
of geometry and composition of the structure under consideration.
However, the qualitative signatures of the proposed mechanism is
expected to persist and may provide a ground for experimental
verification. It should also be pointed out that the framework of
the developed theoretical model allows more accurate estimation
when the detailed information on phonon dispersion and relaxation
is taken into account.


\vspace{0.2in} \noindent {\bf Acknowledgement}

This work was supported in part by the Defense Advanced Research Projects
Agency.

\end{document}